\documentclass[pre,reprint,superscriptaddress]{revtex4}
\pdfoutput=1
\usepackage{graphicx,color,epsfig}
\usepackage{amssymb,amsmath}

\begin{document}
\title{Controlling systemic corruption through group size and salary dispersion \\of public servants}
\author{P. Valverde} 
\affiliation{Pontificia Universidad Cat\'olica del Ecuador, Facultad de Ciencias Exactas y Naturales, Quito, Ecuador}
\author{J. Fernandez}
\affiliation{Pontificia Universidad Cat\'olica del Ecuador, Facultad de Ciencias Exactas y Naturales, Quito, Ecuador}
\author{E. Buena\~no}
\affiliation{Pontificia Universidad Cat\'olica del Ecuador, Facultad de Ciencias Exactas y Naturales, Quito, Ecuador}
\author{J. C. Gonz\'alez-Avella}
\affiliation{Instituto de F\'isica Interdisciplinar y Sistemas Complejos, UIB-CSIC, Palma de Mallorca, Spain}
\affiliation{Advanced Programming Solutions SL, Palma de Mallorca, Spain}
\author{M. G. Cosenza}
\affiliation{Escuela de Ciencias F\'isicas 
y Nanotecnolog\'ia, Universidad Yachay Tech, Urcuqu\'i, Ecuador}

\begin{abstract}
We investigate an agent-based model for the emergence of corruption in public contracts. There are two types of agents: business people and public servants. Both business people and public servants can adopt two strategies: corrupt or honest behavior. Interactions between business people and public servants take place through defined payoff rules. Either type of agent can switch between corrupt or honest strategies by comparing their payoffs after interacting.  We measure the level of corruption in the system by the fractions of corrupt and honest agents 
for asymptotic times. We study the effects of the group size of the interacting agents, the dispersion with respect to the average salary of the public servants, and a parameter representing the institutional control of corruption. We characterize the fractions of honest and corrupt agents as functions of these variables. 
We construct phase diagrams for the level of corruption in the system in terms of these variables, where three collective states can be distinguished: i) a phase 
where corruption dominates; ii) a phase where corruption remains in less than $50\%$ of the agents; and iii) a phase where corruption disappear.
Our results indicate that a combination of large  group sizes of interacting servants and business people and small dispersion of the salaries of public servants, contributes to the decrease of systemic corruption in public contracts.
 \end{abstract}
 \maketitle

\section{\label{sec:level1}Introduction}

Although corruption scandals have increased worldwide in recent years, this is not a new phenomenon as it has been present in all societies and throughout history ~\cite{AbbikSerra2012}. The impact of this problem is not trivial; according to the World Bank \cite{Anderson2019}, 
corruption affects the poorest and most vulnerable population in particular, diverting public resources, slowing economic development and weakening the provision of public services by the government.

Corruption is a phenomenon associated with human behavior about which
it is difficult to obtain reliable observations, either because of its 'illegal' condition whose nature is highly concealed or because of the difficulty in measuring the factors that induce it ~\cite{Theobald1990, Andvig1991}. Therefore, an alternative way for studying this phenomenon is through models and computer simulations, an approach not commonly employed in the social sciences, where conceptual debates that confront approaches, perspectives, and schools, are more prevalent ~\cite{Rodriguez2014}. Corruption should be understood as a multifaceted phenomenon and the results obtained may depend on the method, design and factors included in the model.
 
Game theory and the application of laboratory experiments have made it possible to investigate scenarios where corruption, through bribery, is presented as a dominant strategy ~\cite{Macrae1982}. In these studies, the game is posed through the classic prisoner's dilemma where two players are identified; the possible strategies to execute (to be corrupt or to be honest); and the gains (payoffs) that each player receives for each combination of options that they can choose in the strategic interaction with the other (honest-corrupt; corrupt-corrupt; honest-honest). However, games between two individuals can hardly cover the real spectrum of interactions~\cite{Venkateswaran302265}. 
Other studies have analyzed the role of the salary of public officials, showing no clear evidence in favor of the link between corruption and public sector salaries ~\cite{RAUCH200049, TREISMAN2000399, VANRIJCKEGHEM2001307, DitellaSchargrodsky2003, svensson2005}.

On the control aspect, it has been found that, when private returns to corruption are high or when the possibility or consequences of detection are limited because of weak institutions, individuals are more inclined to act corruptly ~\cite{Abbink2000Fair}.  Similarly, when the economic potential is poor, strong government intervention is more effective in dissuading free riders, provided that this intervention is combined with other strategies to mitigate corruption.

Some models have 
 studied the behavior of actors in the face of corruption through social intimidation~\cite{Bauza2020}. They find that, if social intimidation is weak, there are three equilibrium states: complete honesty, full corruption and a mixed state, which are connected through smooth transitions. However, when social intimidation is strong, the transitions become explosive and lead to a bistable phase where complete corruption and full honesty coexist. Weisel et al. ~\cite{Weisel2015} conducted an experiment with students to explore the collaborative forces behind corruption. Their findings reaffirm the existence of a functional morality and reveal a dark side of collaboration: when participants are faced with opposing moral sentiments (being honest versus joining forces in collaboration) people choose to collaborate in an act of corruption so as not to see their benefit diminished compared to others.

In this article, we  propose an 
agent-based model of a public good game 
with two types of participants
to simulate the decision process for honest or corrupt behavior by contractors and officials in a public procurement. The model includes several relevant elements that have been considered separately, such as the  number of actors involved in the interactions, the level of control, and the dispersion of the salaries received by the officials. Agents can change their behavior between honest or corrupt depending on their incentives and their local interactions.
Our model allows to observe different strategies or combinations of variables that could be adopted by the State in order to achieve institutions with lower levels of corruption without increasing the levels of control.

\section{Game model for Public Corruption}
We assume the existence of two types of agents in the system: public servants and contractors or business-people.
Both can adopt, in each iteration of the game, one of two strategies: honest (H) or corrupt behavior (C). 
Interactions occur between public servants and business-people through some payoff rules. Agents can maintain or change their strategy by observing their payments in previous iterations of the game and those of the agents with whom they are linked in their interaction group.
In Table~1 we define the payoff matrix that assigns the payments to the agents in each public servant-business person interaction according to their strategies \cite{wydick_2007}.

 \begin{table}[h]
 	\centering
 	\includegraphics[scale=2.5]{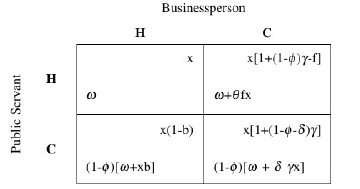}
  \caption{Payoff matrix.}
 \end{table}	

On each box, the upper expression corresponds to the payment received by the business person, and the lower expression indicates the payment going to the public servant. 
The parameters appearing in the payoff matrix are the following: $x$ is the real value of the public contract; $\omega$ is the public servant wage; $f$ is the fine imposed by the honest public servant to the corrupt business person;  $b$ is the bribery share that the corrupt servant receives; $\theta$ is the fraction of the fine applied to the corrupt business person that the public servant receives as an incentive for detecting corruption; $\gamma$ is the fraction of the public contract surcharge; $\delta$ is the surcharge fraction received by the corrupt servant; $\phi$ is a Bernoulli random variable with parameter $c$; and $c$ is the probability that an external institutional control detects corrupt behavior in any agent.

\subsection{Interaction dynamics for the model}
We consider a system of $N$ agents divided into two populations: a population of $N/2$ public servants and a
population of $N/2$ business-people or contractors.  Each agent in either population can adopt one of two possible strategies: C (corrupt) and H (honest). The strategies are initially assigned to the agents in the system at random with a uniform distribution. Therefore, there are on average $N/2$ agents on each strategy in the initial system. The salary of each public servant ($\omega$) is initially assigned at random according to a log-normal distribution, where $\bar{\omega}$ is the mean of the salary and $\sigma$ is the standard deviation of such distribution.  The parameter $\sigma$ represents the  heterogeneity of
the salaries of the public servants.
Then, within the population of public servant, $N/2$ subsets of equal size $g_s$ are generated. Similarly, the population of contractors is divided into subsets of equal size $g_s$.
In one realization of the dynamics, these subsets are generated as follows. For each element within a
population, called the central agent $i$, we select at random
$g_s-1$ neighbors in that population.

Then, given the parameters defined in the payoff matrix, the dynamics of the model is defined as follows.

\begin{enumerate}
\item At one iteration of the game, 
  each public servant in sequence interacts with a randomly chosen business person according to the rules of the payoff matrix.
\item After $N/2$ iterations, 
all agents collect their profits and their strategies are synchronously updated. For each subset $g_s$, the central agent $i$ chooses  randomly one neighbor $j$ and compares their respective payoffs $p_i$ and $p_j$: if $p_i < p_j$,  agent $i$ adopts the strategy of agent $j$, otherwise agent $i$ remains in its strategy.
\item After $N/2$ iterations the time step increases by one.
\end{enumerate}

\section{Results and Discussion}

\subsection{Collective behavior on the space of parameters $(c,\sigma)$}

In order to investigate the collective behavior of the
system, we calculate the fraction (or density) of public servants that share the honest state $H$, denoted by $\rho_{poh}$. Similarly, we measure the density of busines people in the honest state $H$, called $\rho_{bh}$.
The density of honest agents $H$ in the system, including public servants and contractors, is denoted by $\rho_{h}$, while the density of corrupt agents $C$ in the system is indicated by $\rho_{c}$.

\begin{figure}[h]
	\centering
\includegraphics[scale=0.8]{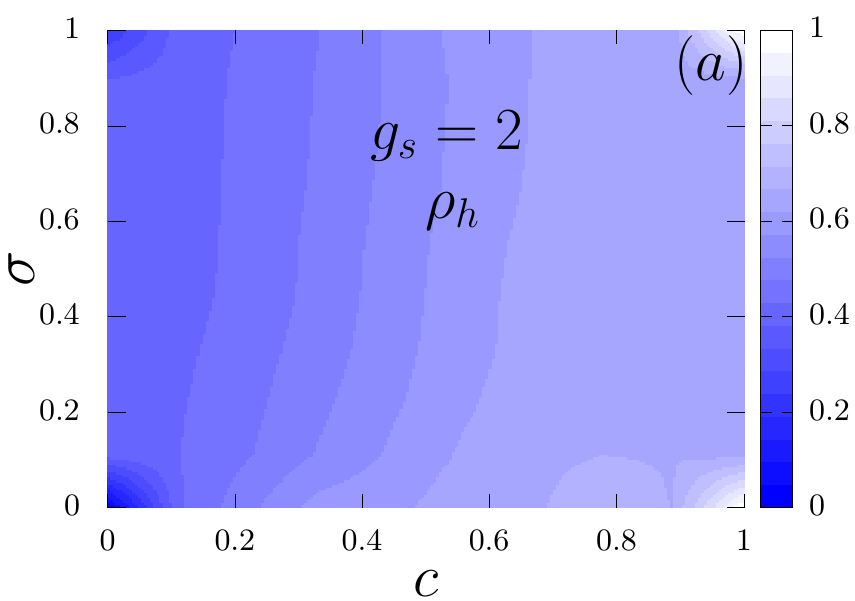}
\hspace{0.1cm}
\includegraphics[scale=0.8]{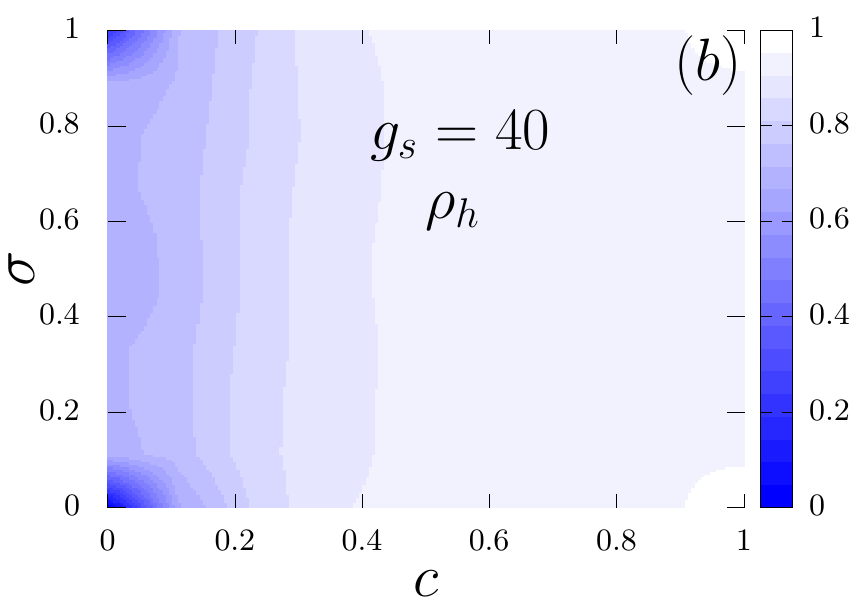}\\
\includegraphics[scale=0.8]{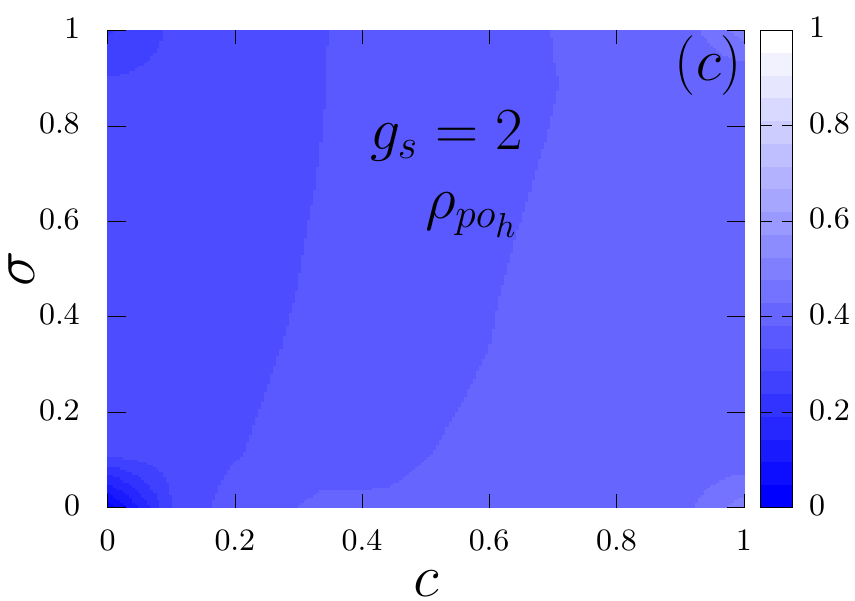}
\hspace{0.1cm}
\includegraphics[scale=0.8]{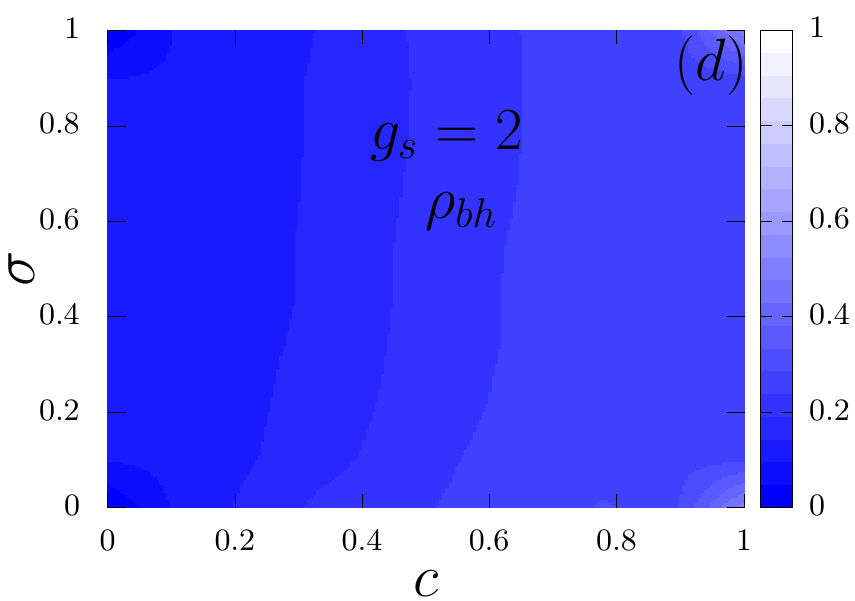}
   \caption{\label{fig:fig1}
 Density or fraction of honest agents, indicated by a color code on the right of each panel, on the space of parameters given by the standard deviation of salaries $\sigma$ and the corruption control probability $c$. a) Density of honest agents in the system  $\rho_{h}$, including public servants and contractors, for group size $g_s=2$. b) Density of honest agents in the system  $\rho_{h}$, for group size $g_s = 40$. 
 c)  Density of honest public servants $\rho_{poh}$, for group size $g_s=2$. d) Density of honest business-people $\rho_{bh}$,
 for group size $g_s=2$. Fixed parameters are:  system size $N = 1000$, mean salary of public servants $\bar\omega= 0.5$, $\gamma = 0.1$, $\delta = 0.1$, $\theta =0.01$, $f = 0.1$, $x = 1.0$ (normalized real value of public contract), and 
 $b = 0.1$.}
 \end{figure}

In Figure \ref{fig:fig1} we show these densities, indicated by a color code, as functions of two parameters:
the corruption control probability $c$ and the standard deviation of salaries $\sigma$.
Figures~\ref{fig:fig1}(a) and \ref{fig:fig1}(b) show $\rho_{h}$ on the space of parameters $(c,\sigma)$ for two different values of group sizes
$g_s$.  We observe in Fig.~\ref{fig:fig1}(a) that, when the group size is small ($g_s = 2$), the region of parameters leading to honest behavior characterized by $\rho_{h}\geq 0.5$ is smaller than this corresponding region for larger group size ($g_s=40$), as seen in Fig.~\ref{fig:fig1}(b).
It can be noticed that, as the corruption control probability $c$ increases, the region of parameters for honest agents $\rho_{h}\geq 0.5$ also increases, for both small and large groups. The dispersion of salaries has less influence on the honest behavior in the system, for a given level of corruption control. We may expect that, for small groups as in Fig.~\ref{fig:fig1}(a), the incidence of corruption is higher because the agents know the payoff of their neighbors, and the most likely adopted strategy is that of corrupt when corruption control levels are low. On the other hand, in large groups, as in Fig.~\ref{fig:fig1}(b), it is more difficult to know the strategies adopted by the neighbors, and there should be greater dependence on the control of corruption.

To compare the relative proportions of corruption for public servants and contractors, Fig. \ref{fig:fig1}(c) and Fig. \ref{fig:fig1}(d) show their respective densities $\rho_{poh}$ and $\rho_{bh}$ on the plane $(c,\sigma)$ with the same group size. We observe that the region of prevalence of honest behavior for business people on the plane $(c,\sigma)$ characterized by $\rho_{bh}\geq 0.5$ is much smaller than the corresponding region of honest behavior for 
public servants with $\rho_{poh} \geq 0.5$.  Our model indicates that corruption is more extended among business people than among public servants. 
The business person or contractor is the one who determines the overprice and, if he/she is not detected, he/she receives the largest proportion of that amount. Therefore, he/she has a greater propensity to behave in a corrupt manner, compared to the public servant who receives a fraction of that overprice, when not detected. However, when the group size increases, honest behavior increases for both types of agents.

\subsection{Relation between the density of corrupt agents and the control of corruption}

\begin{figure}[h]
	\centering
	\includegraphics[scale=0.4]{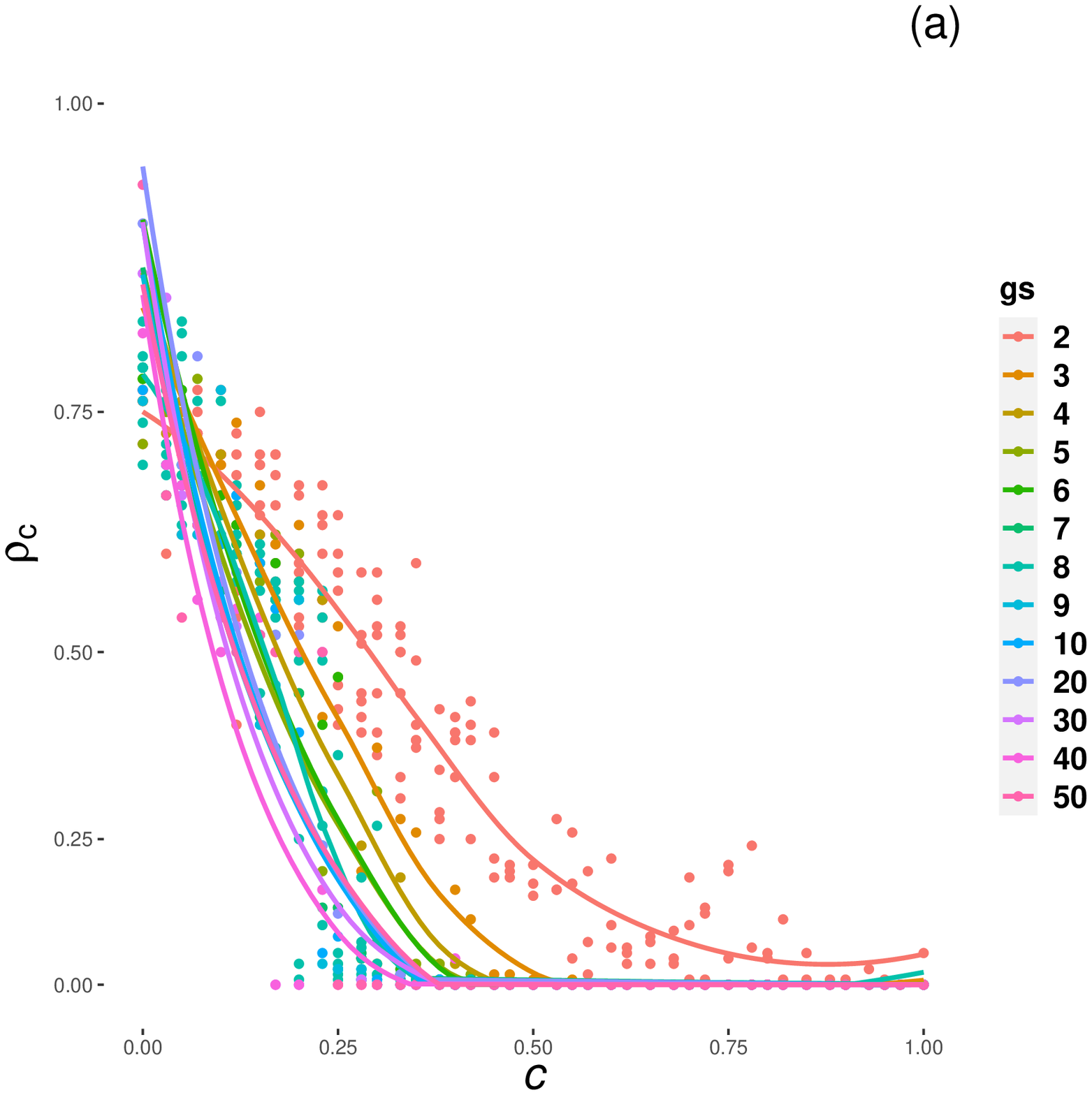}
\hspace{4mm}
\includegraphics[scale=0.4]{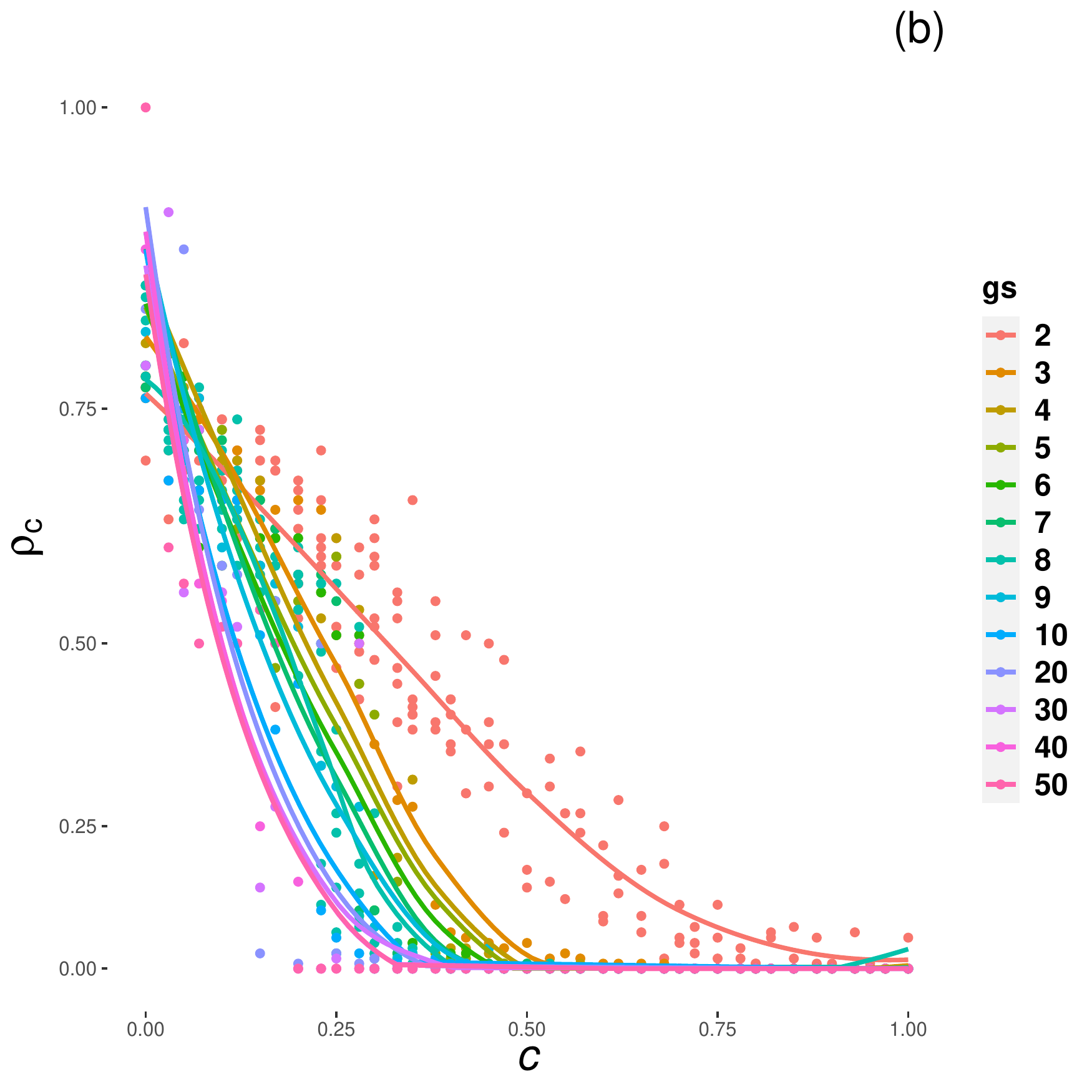}\\
\includegraphics[scale=0.4]{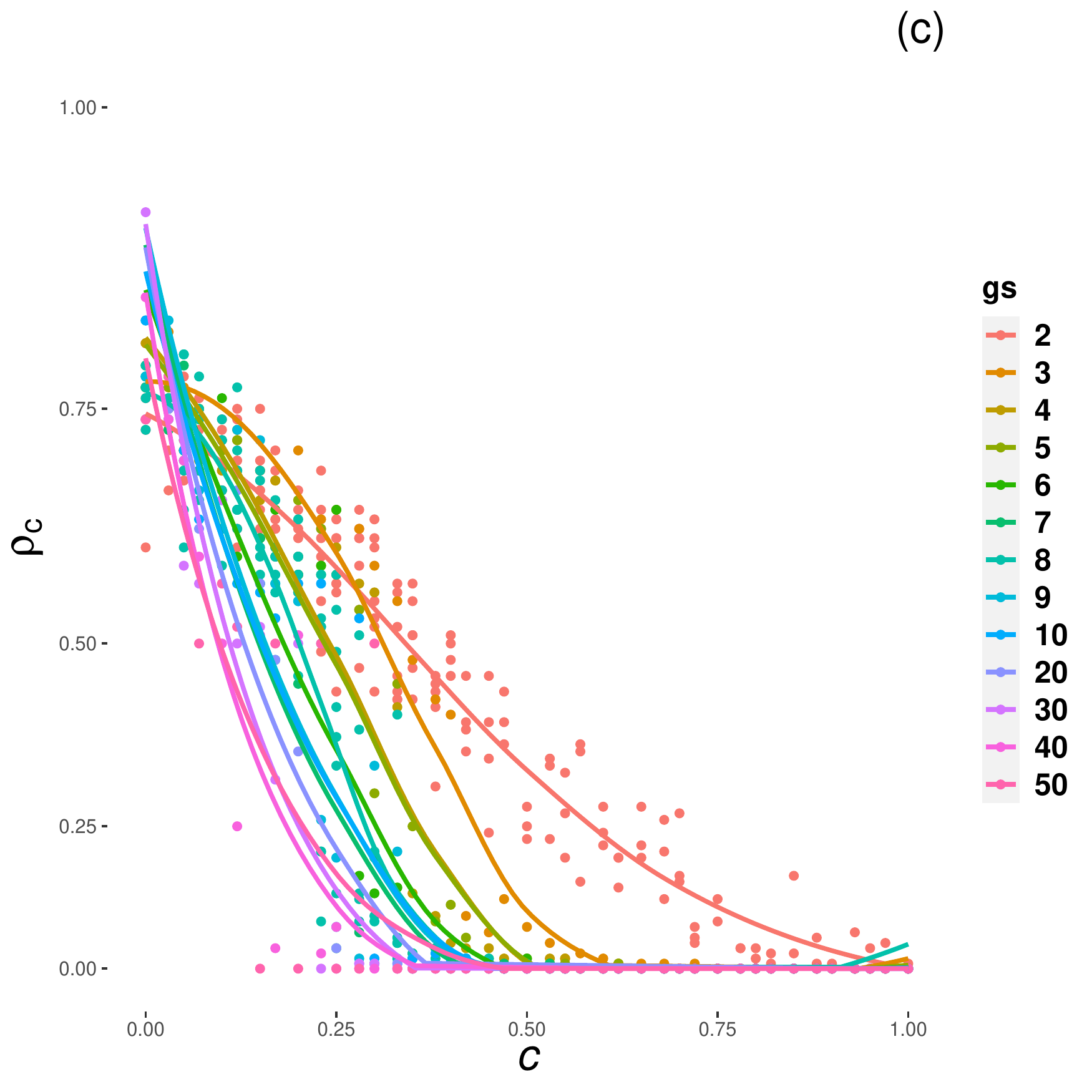}
\caption{Density of corrupt agents $\rho_c$ as a function of the corruption control probability $c$, for various  group sizes $g_s$ and different values of the standard deviations of the salaries $\sigma$.
Fitting curves are displayed for each value of $g_s$. a) $\sigma= 0.1$.  b) $\sigma = 0.25$. c) $\sigma = 0.5$.
Fixed parameters: system size $N = 1000$, $\bar\omega= 0.5$, $\gamma = 0.1$, $\delta = 0.1$, $\theta =0.01$, $f = 0.1$, $x = 1.0$ (normalized real value of public contract), and $b = 0.1$.}
\label{fig:fig2}
\end{figure}

To investigate the influence of the salary dispersion of public servants $\sigma$ and the group size in the emergence of corruption, in Fig.~\ref{fig:fig2} we show the fraction of corrupt agents in the system, defined as $\rho_c=1-\rho_h$, as a function of the corruption control probability $c$ for different values of $\sigma$ and group sizes $g_s$. The three panels in Fig.~\ref{fig:fig2} correspond to the values: (a) $\sigma= 0.1$,  (b) $\sigma = 0.25$, and (c) $\sigma = 0.5$.

When the size of the groups is large, the functional relation between  $\rho_c$ and
$c$ appears independent of the dispersion $\sigma$, as the fitting curve for $g_s=50$ shows in the three panels of Fig.~\ref{fig:fig2}. For large groups, the dispersion of salaries has practically no effect in the level of corruption in the system. For small group sizes, the dispersion of salaries plays a role in
the density of corruption. For example, the fitting curve for $g_s=2$ in Fig.~\ref{fig:fig2}(a) decays faster than the curve for $g_s=2$ in Fig.~\ref{fig:fig2}(c). 
As the dispersion of salaries increases, corruption is more persistent in the presence of small size groups; there is a greater propensity to adopt the corrupt behavior strategy which can bring higher earnings.

\subsection{Critical values for control of corruption}
\begin{figure}[h]
  \centering
\includegraphics[scale=0.6]{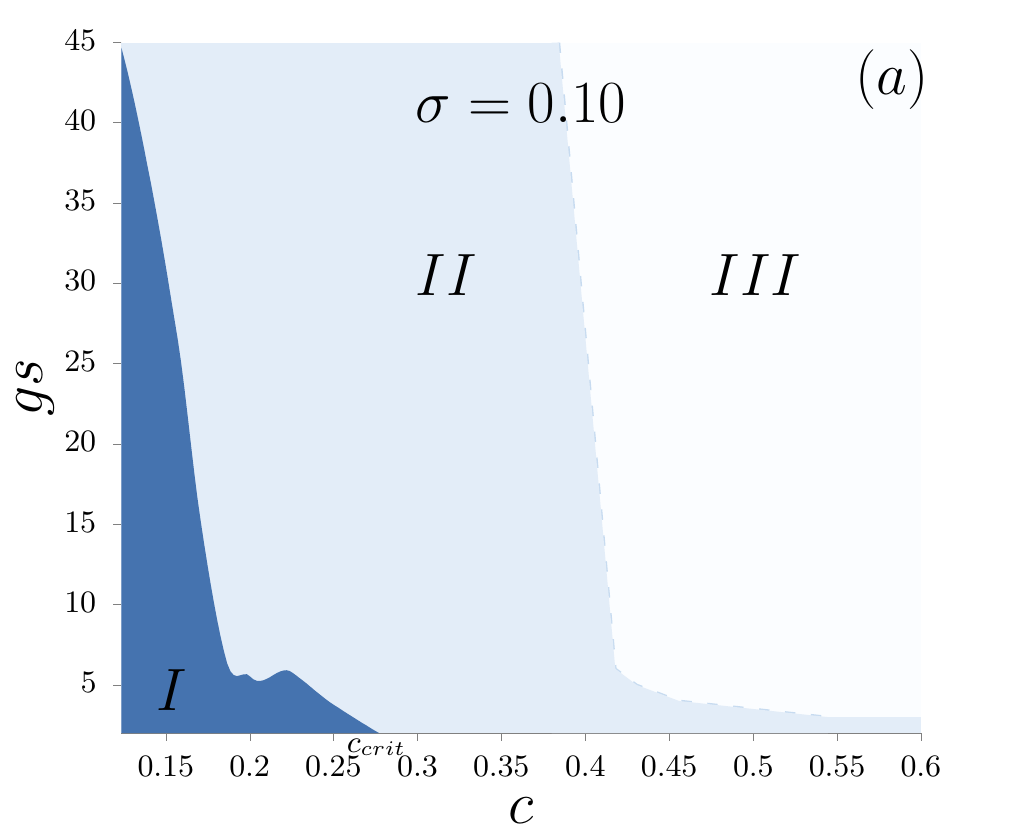}
\hspace{4mm}
\includegraphics[scale=0.6]{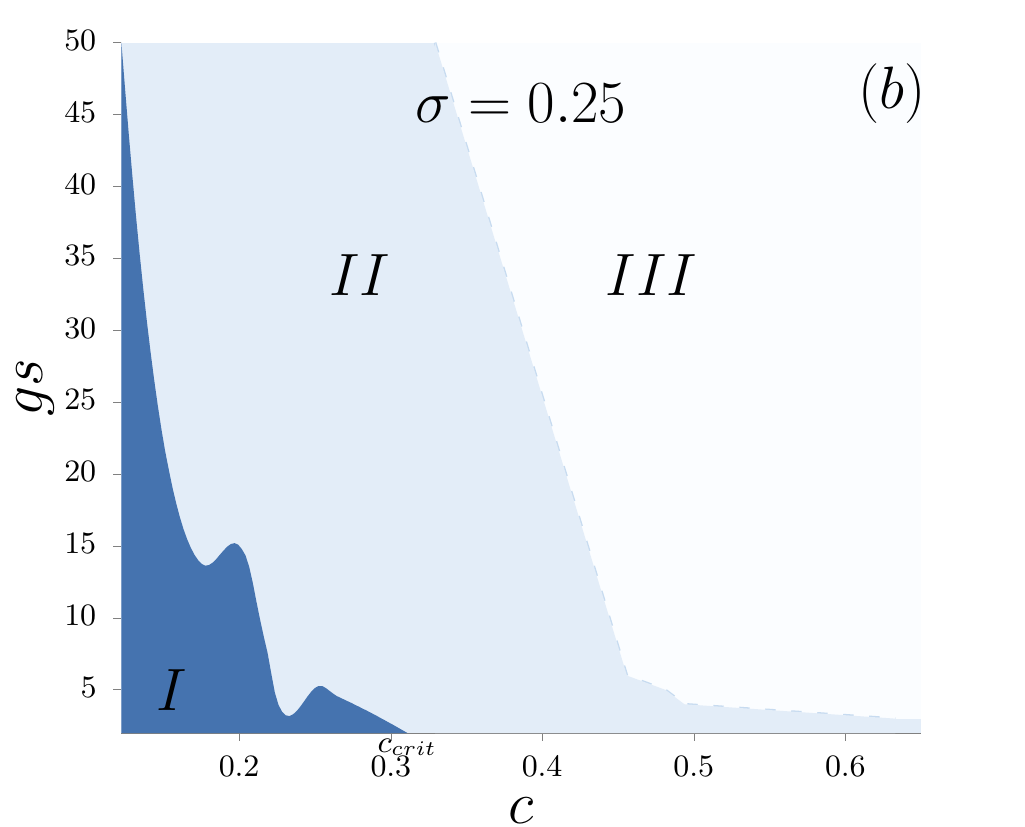}\\
\includegraphics[scale=0.6]{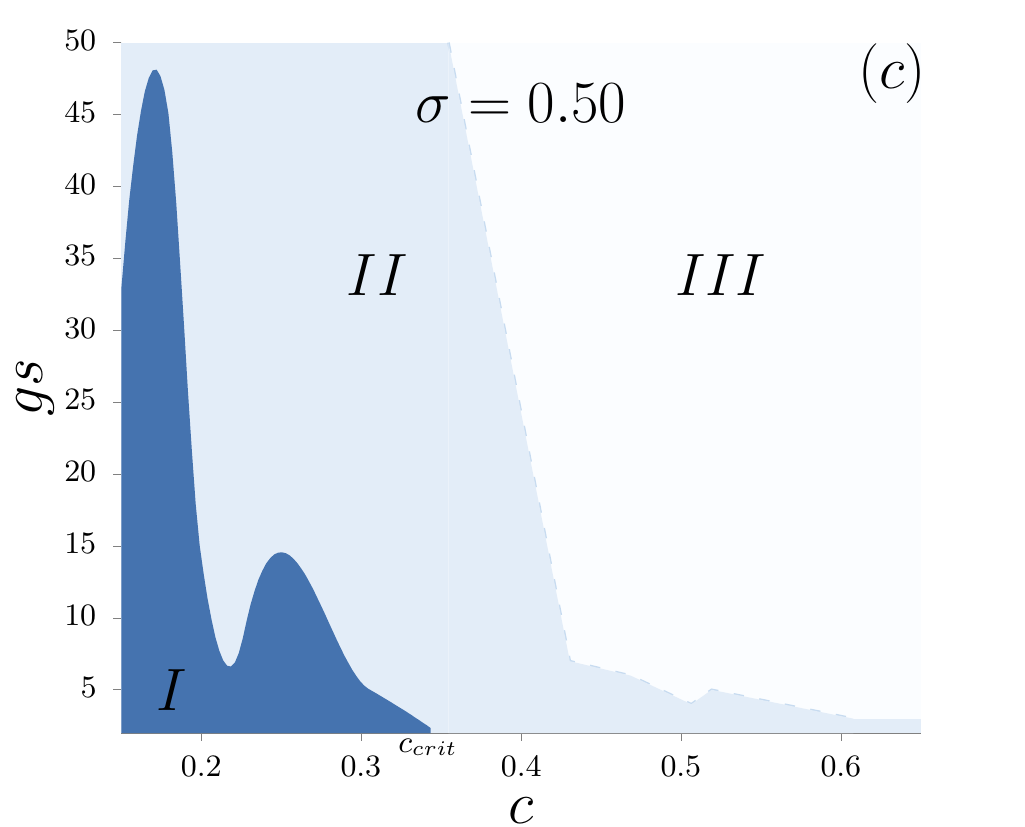}
   \caption{Critical curves on the space of parameters $(g_s,c)$ for which  $\rho_c=0.5$ and $\rho=0$, for different values of the dispersion of salaries $\sigma$. On each panel, three regions of collective behaviors or phases are indicated with different shades of color: I) systematic corruption, characterized by values $\rho_c > 50\%$ with dark color; II) controlled corruption, characterized by values
   	$\rho_c < 50\%$ with light color; and III) no corruption or honesty zone for which $\rho_c =0$ with white color. 
   	a) $\sigma = 0.10$. b) $\sigma = 0.25$. c) $\sigma = 0.50$.}
\label{fig:fig3}
\end{figure}

From Fig.~\ref{fig:fig2} we can obtain, for a given value of $\sigma$, the values of the group size $g_s$ and the corruption control parameter $c$ for which the density of corrupt agents in the system has a specific value. In Fig.~\ref{fig:fig3} we show the resulting functional relations $g_s$ versus $c$ corresponding to the values $\rho_c=0.5$ and $\rho=0$.
For each value of $\sigma$, the relation  $g_s$ versus $c$ defines critical curves on the space of parameters $(g_s,c)$ that separates
regions parameters where distinct behaviors of the system arise. We identify three collective behaviors or phases from Fig.~\ref{fig:fig3}: I) systematic corruption, characterized by values $\rho_c > 50\%$ below the critical curve; II) controlled corruption, characterized by values
$\rho_c < 50\%$; and III) no corruption or honesty zone where $\rho_c =0$.
 Figure~\ref{fig:fig3} reveals that, for a given dispersion of salaries, a combination of 
 external control and group sizes can lead to a prescribed level of corruption in the system. In particular, corruption can be decreased by reducing the sizes of the groups of interacting agents, which is usually less expensive than increasing the mechanisms of control of the corruption. On the other hand, institutional control of corruption above some critical level can extinguish the corrupt behavior of the agents.

\subsection{Salary dispersion and critical control of systematic corruption}
\begin{figure}[h]
\centering
\includegraphics[scale=0.6]{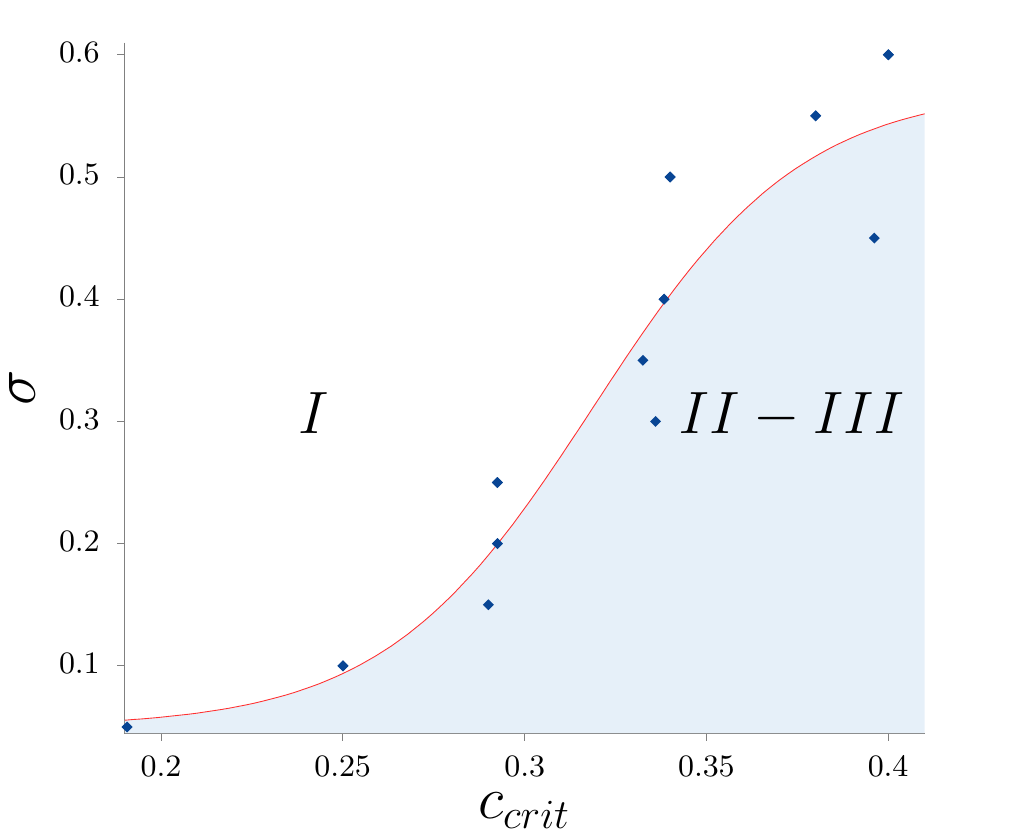}
\caption{Dispersion of salaries $\sigma$ as a function of the critical values of the corruption control parameter $c_{crit}$ for the onset the majority honesty zone II. The numerically calculated data and the corresponding fitting curve are shown. The curve separates two regions: i) a region where phase I  exist; and ii) a region denoted below the curve corresponding to phases II and III. Fixed parameters: $N = 1000$,  $\bar\omega= 0.5$, $\gamma = 0.1$, $\delta = 0.1$, $\theta =0.01$, $f = 0.1$, $x = 1.0$, and $b = 0.1$}
\label{fig:fig4}
\end{figure}

Figure~\ref{fig:fig3} indicates that, for each value of the dispersion of salaries $\sigma$, there is a maximum value of the corruption control probability $c_{crit}$ for the onset of phase II where a majority ($\rho_c<0.5$) of honest public servants exist. The value $c_{crit}$ corresponds to the minimum group size $g_s=2$. 
 Thus, we have calculated the dependence of the $c_{crit}$ values on the dispersion of salaries $\sigma$. The numerically calculated data and the corresponding fitting curve are shown in Figure~\ref{fig:fig4}.
The dispersion of salaries tends to a constant value as the control parameter increases.
The curve $\sigma$ versus $c_{crit}$ separates two regions with distinct collective behaviors: i) a region above the curve where phase I can be found, with high level of corruption in the system; and ii) a region below the curve corresponding to the majority honest behavior zone of phases II and III.
Figure~\ref{fig:fig4} signals that there exists a maximum value of the dispersion of salaries above which a dominant honest behavior cannot be achieved by institutional control of corruption. 

\section{Conclusions}
Corruption in public contracts is a complex phenomenon associated with human behavior that depends on multiple factors. In this article we have investigated
an agent-based model of a public good game involving several variables with two types of participants:  public servants and business
people. Participants can change their strategies between honest or corrupt, depending on their local interactions and their incentives.

Although the emergence of public corruption can be seen from diverse perspectives, have focused on the collective
behavior associated to three parameters in our model: the standard deviation or dispersion of the salaries, the size of the interaction groups, and the probability of institutional control.
Our results show that there are ranges of these parameters where the emergence of corrupt behavior in public contracts can be reduced and even can be made to disappear.

A small salary dispersion together with small institutional control of corruption can keep a low level of corrupt behavior in the system. 
The group size also plays an important role for the presence of corruption: smaller interaction groups of agents tend to favor the
the appearance of corrupt behavior.  
Our model also indicates that, since no penalty condition is applied to corrupt behavior of contractors, these type of agents become the most corrupt.

Control of corruption in public institutions requires effort and resources \cite{VALVERDE201761}. Our results suggest measures that are relatively simple and have low cost to implement.
Future research should explore models with alternative methods for controlling public corruption; for example a decentralized system of control such as the blockchain system \cite{nakamoto2009bitcoin}.

\section*{Acknowledgments}
This research was supported by Corporaci\'on Ecuatoriana para el Desarrollo de la Investigaci\'on y Academia (CEDIA) through project CEPRA XVI-2022-09 ``Desarrollo y Aplicaciones de Recursos Computacionales en Sociof\'isica''. We are grateful to the Research Department of PUCE, Ecuador, for support.

\newpage

\bibliographystyle{elsarticle-num-names}
\nocite{*}
\bibliography{main}

\end{document}